\documentclass[10pt,a4paper,conference]{IEEEtran}
\usepackage[latin9]{inputenc}
\usepackage{units}
\usepackage{bm}
\usepackage{amsthm}
\usepackage{amsmath}
\usepackage{amssymb}
\usepackage{graphicx}

\makeatletter

\pdfpageheight\paperheight
\pdfpagewidth\paperwidth

\providecommand{\tabularnewline}{\\}

\theoremstyle{plain}
\newtheorem{thm}{\protect\theoremname}
\theoremstyle{remark}
\newtheorem{rem}[thm]{\protect\remarkname}

\usepackage{cite}
\usepackage{units}
\usepackage{txfonts}

\newcommand{\unfootnote}[1]{
  \renewcommand{\@makefnmark}{}
  \footnotetext{#1}
  \renewcommand{\@makefnmark}{\mbox{$^{\@thefnmark}$}}
}

\usepackage{skak}

\makeatother

\providecommand{\remarkname}{Remark}
\providecommand{\theoremname}{Theorem}

\begin{document}

\title{Ordered Tomlinson-Harashima Precoding\\
 in G.fast Downstream}

\author{\IEEEauthorblockN{Miroslav Hekrdla\IEEEauthorrefmark{1}, Andrea Matera\IEEEauthorrefmark{1}, Weiyang Wang\IEEEauthorrefmark{2}, Dong Wei\IEEEauthorrefmark{2} and Umberto Spagnolini\IEEEauthorrefmark{1}}

\IEEEauthorblockA{\IEEEauthorrefmark{1}Dipartimento di Elettronica, Informazione e Bioingegneria (DEIB), Politecnico di Milano, Italy}

\IEEEauthorblockA{\IEEEauthorrefmark{2}Huawei Technology Co., Ltd., Shenzhen, Peoples Republic of China\\
E-mails: \IEEEauthorrefmark{1}\{miroslav.hekrdla, andrea.matera, umberto.spagnolini\}@polimi.it, \IEEEauthorrefmark{2}\{wangweiyang.wang, weidong\}@huawei.com}}
\maketitle
\begin{abstract}
G.fast is an upcoming next generation DSL standard envisioned to use
bandwidth up to 212 MHz. Far-end crosstalk (FEXT) at these frequencies
greatly overcomes direct links. Its cancellation based on non-linear
Tomlinson-Harashima Precoding (THP) proved to show significant advantage
over standard linear precoding. This paper proposes a novel THP structure
in which ordering of successive interference pre-cancellation can
be optimized for downstream with non-cooperating receivers. The optimized
scheme is compared to existing THP structure denoted as equal-rate
THP which is widely adopted in wireless downlink. Structure and performance
of both methods differ significantly favoring the proposed scheme.
The ordering that maximizes the minimum rate (max-min fairness) for
each tone of the discrete multi-tone modulation is the familiar V-BLAST
ordering. However, V-BLAST does not lead to the global maximum when
applied independently on each tone. The proposed novel Dynamic Ordering
(DO) strategy takes into account asymmetric channel statistics to
yield the highest minimum aggregated rate.\unfootnote{{\it Accepted at the 2015 IEEE Globecom 2015, Selected Areas in Communications: Access Networks and Systems}, 6-10 December, 2015.}
\end{abstract}

\section{Introduction}

\begin{figure}[h]
\begin{centering}
\includegraphics[width=0.8\columnwidth]{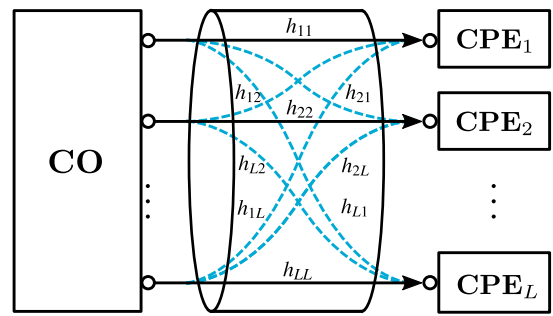}
\par\end{centering}

\protect\caption{Downstream FEXT channel model.\label{fig:Downstream-FEXT-channel}}
\end{figure}

Digital Subscriber Line (DSL) is dominant broadband access technology
due to its ability to fulfill demands for reliable high-data-rate
connectivity in a cost-effective way by exploiting the existing infrastructure
of twisted-pair copper lines. Upcoming next generation DSL standard,
G.fast~\cite{Timmers-Guenach-etal_2013}, proceeds in the trend of
shortening copper lines (up to 250\,m) between Central Office (CO)
and Consumer Premised Equipment (CPE) aiming at fiber-like connection
(up to 1\,Gbps). Short lines enable the usage of wider bandwidth
(initially up to 106\,MHz extended later to 212\,MHz) than used
by the foregoing Very-high-bit-rate DSL (VDSL2) standard operating
up to 30\,MHz. Cancellation of crosstalk between the lines by multi-user
processing (denoted as vectoring or signal coordination) has a major
impact on system performance. Far-End-Crosstalk (FEXT) is the crosstalk
affecting the other end of the line w.r.t. the transmitter as shown
for Downstream (DS) in Fig.\,\ref{fig:Downstream-FEXT-channel}.
FEXT is typically canceled by suitable transmitter precoding. If signal
coordination is restricted (e.g. multiple non-cooperating providers
in the same cable bundle), spectrum coordination (dynamic spectrum
management) is applied~\cite{Huberman-Leung-Ngoc_2012}. Diagonal
Precoding (DP) is a linear FEXT cancellation precoding adopted in
VDSL2. It performs at the information theoretical limits in VDSL2
band where FEXT channel is much weaker than direct lines~\cite{Cendrillon-Moonen-etal_2004}.
Non-linear FEXT cancellation based on Tomlinson-Harashima Precoding
(THP)~\cite{Ginis-Cioffi_2000} provides significant gains over linear
precoding in G.fast band where FEXT is often as strong as direct lines~\cite{Muller-Lu-etal_2014}. 

This paper modifies the THP proposed in~\cite{Ginis-Cioffi_2000}
by introducing an ordering of successive interference pre-cancellation
performed by CO that is  optimized for non-cooperating CPEs. Ordering
optimization has been already presented in~\cite{Fischer-Windpassinger-etal_2002},~\cite{Joham-Brehmer-Utschick_2004}
for a different THP structure which we denote as Equal-Rate THP (ER-THP).
ER-THP provides constant Signal-to-Noise Ratio (SNR) at each line.
Any type of ordering can be concatenated with the proposed THP scheme.
The ordering which maximizes the minimum rate on a single tone of
Discrete Multi-Tone (DMT) modulation is V-BLAST (VB) ordering~\cite{Wolniansky-Foschini-etal_1998},~\cite{Wubben-Bohnke-etal_2001}.
However, VB does not provide the maximal minimum of aggregated rates
when applied on all DMT tones. We propose a novel Dynamic Ordering
(DO) strategy that takes into account the asymmetry of G.fast channel
statistics. The proposed scheme together with DO provides the maximal
minimum rate of $\sim955$\,Mbps over tested 100\,m long paper-insulated
cable. The ordering arbitrary adjusting general user demands require
considerable computation power~\cite{Kerpez-Ginis_2014}, complexity
of DO is from this perspective negligible.

\emph{Notation: }Bold upper- and lower-case letters describe matrices
and column vectors. $\left[\mathbf{A}\right]_{ij}=a_{ij}$ denotes
the \emph{ij}th element of matrix $\mathbf{A}.$ Letters $\mathbb{Z},\mathbb{Z}_{\mathrm{j}},\mathbb{R},\mathbb{C}$
refer to integers, complex integers, real and complex numbers, respectively.
We denote matrix inversion, transposition and conjugate transposition
as $\left(\star\right)^{-1},\left(\star\right)^{T},\left(\star\right)^{H}$.
Symbol $\triangleq$ denotes equality from definition.

\emph{Organization: }Section~\ref{sec:System-Model} describes general
THP scheme and its properties in which reference method~\cite{Ginis-Cioffi_2000}
is defined. The proposed scheme is described in Sec.\,\ref{sec:Proposed-Ordered-THP}
and some ordering strategies are in Sec.\,\ref{sec:Optimized-Ordering-of}.
Comparison to ER-THP, numerical results and conclusions are content
of Sec.\,\ref{sec:Comparison-with-Ordered}, \ref{sec:Performance-Evaluation}
and \ref{sec:Conclusion}.

\section{System Model\label{sec:System-Model}}

\subsection{Downstream Channel Model}

We assume centralized transmission from CO to non-cooperating CPEs
in DS as shown in Fig.\,\ref{fig:Downstream-FEXT-channel}. DMT is
employed to turn the frequency selective channel into a set of frequency
flat orthogonal channels. On each tone, the signal received by $L$
CPEs $\mathbf{y}\in\mathbb{C}^{L}$ is modeled as 
\begin{equation}
\mathbf{y}=\mathbf{H}\mathbf{x}+\mathbf{w},
\end{equation}
where $\mathbf{x}=[x_{1},\dots,x_{L}]^{T}\in\mathbb{C}^{L}$ denotes
the transmitted signal vector and $\mathbf{w}\in\mathbb{C}^{L}$ is
the AWGN. The cable bundle is assumed to contain only the $L$ lines.
We avoid tone indexing to simplify the notation. Main diagonal elements
$h_{ii}=\left[\mathbf{H}\right]_{ii}$ of channel matrix $\mathbf{H}\in\mathbb{C}^{L\times L}$
characterize insertion loss of direct lines, and off-diagonal elements
$h_{ij}=\left[\mathbf{H}\right]_{ij}$ with $i\not=j$ characterize
FEXT. The channel is static and assumed to be known at the transmitter.

\subsection{General THP Scheme and Basic Properties\label{sub:General-Scheme-of}}

\begin{figure}
\begin{centering}
\includegraphics[width=0.85\columnwidth]{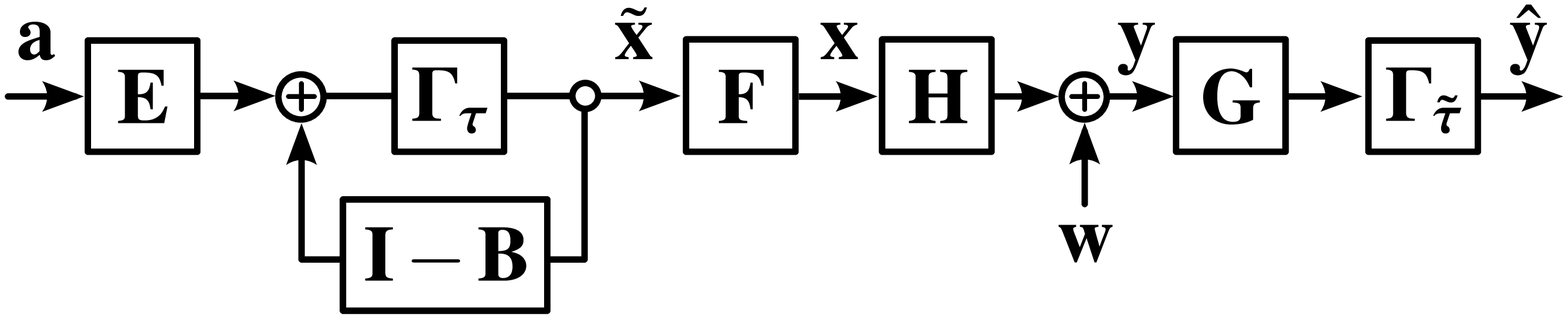}\vspace{-7pt}

\par\end{centering}

\protect\caption{General Tomlinson-Harashima Precoding scheme.\label{fig:General-Tomlinson-Harashima-Prec}}
\vspace{-7pt}
\end{figure}

We describe considered THP schemes in the common framework in Fig.~\ref{fig:General-Tomlinson-Harashima-Prec}.
Linear block $\mathbf{E}$ represents the ordering (or later assumed
lattice reduction). The feedback loop consisting of non-linear modulo
$\bm{\Gamma}_{\bm{\tau}}$ block and linear block given by lower triangular
matrix $\mathbf{B}$ with units along the main diagonal implements
the inversion of \textbf{$\mathbf{B}$ }while reducing transmitted
power by modulo $\bm{\Gamma}_{\bm{\tau}}.$ $\mathbf{F}$ is a feedforward
filter which also ensures transmitted signal to satisfy energy constraints.
Diagonal matrix $\mathbf{G}$ describes linear operations performed
by non-cooperating receivers. Let the input to the precoding chain
be vector $\mathbf{a}=[a_{1},\dots,a_{L}]^{T}\in\mathbb{C}^{L}$ forming
data symbols and the output be decision variable vector $\hat{\mathbf{y}}\in\mathbb{C}^{L}.$

\subsubsection{Linearized Scheme}

\begin{figure}
\begin{centering}
\includegraphics[width=0.85\columnwidth]{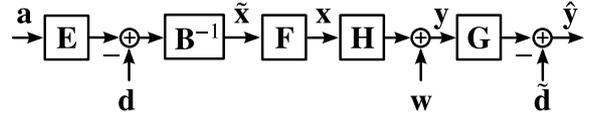}\vspace{-7pt}

\par\end{centering}

\protect\caption{Linearized general THP scheme.\label{fig:Linearized-general-THP}}
\vspace{-7pt}
\end{figure}

Block $\bm{\Gamma}_{\bm{\tau}}$ is the modulo function over base
$\tau$ with origin shifted by $\nicefrac{\tau}{2}$ applied individually
along each dimension of the input $\mathbf{x}.$ Particularly,
\begin{align*}
\Gamma_{\tau}[x] & \triangleq\left(x+\nicefrac{\tau}{2}\right)_{\mathrm{mod}\tau}-\nicefrac{\tau}{2},\quad x\in\mathbb{R},\\
\Gamma_{\tau}[x] & \triangleq\Gamma_{\tau}[\Re\{x\}]+\mathrm{j}\Gamma_{\tau}[\Im\{x\}],\quad x\in\mathbb{C},\\
\bm{\Gamma}_{\bm{\tau}}[\mathbf{x}] & \triangleq\left[\Gamma_{\tau_{1}}[x_{1}],\dots,\Gamma_{\tau_{L}}[x_{2}]\right]^{T},\quad\mathbf{x}\in\mathbb{C}^{L}.
\end{align*}
Every modulo reminder equals to the input minus an integer multiple
of base $\tau$ such that the reminder is lower than $\tau.$ Therefore
$\bm{\Gamma}_{\bm{\tau}}[\mathbf{x}]=\mathbf{x}-\mathbf{d},$ where
$\mathbf{d}$ is a vector such that \emph{$\mathbf{x}-\mathbf{d}\in\left[-\nicefrac{\tau_{1}}{2},\nicefrac{\tau_{1}}{2}\right)\times\dots\times\left[-\nicefrac{\tau_{L}}{2},\nicefrac{\tau_{L}}{2}\right).$}
The \emph{i}th component of $\mathbf{d}$ is $d_{i}\in\tau_{i}\mathbb{Z}_{\mathrm{j}}$
where $\mathbb{Z}_{\mathrm{j}}\triangleq\mathbb{Z}+\mathrm{j}\mathbb{Z}$
denotes complex integers and $\bm{\tau}=[\tau_{1},\dots,\tau_{L}]^{T}$
is a vector of thresholds. Figure~\ref{fig:Linearized-general-THP}
shows the linearized scheme where $\bm{\Gamma}_{\bm{\tau}}$ is replaced
by additive term $-\mathbf{d}.$

\subsubsection{Zero-Forcing Condition}

Zero-Forcing (ZF) precoding inverts the channel by eliminating the
crosstalk such that 
\begin{equation}
\hat{\mathbf{y}}=\mathbf{G}\mathbf{H}\mathbf{F}\mathbf{B}^{-1}\mathbf{E}\mathbf{a}-\mathbf{G}\mathbf{H}\mathbf{F}\mathbf{B}^{-1}\mathbf{d}+\mathbf{G}\mathbf{w}-\tilde{\mathbf{d}}
\end{equation}
 equals to input data $\mathbf{a}$ plus noise. The whole chain of
linear blocks in Fig.\,\ref{fig:Linearized-general-THP} needs to
fulfill ZF condition 
\begin{equation}
\mathbf{G}\mathbf{H}\mathbf{F}\mathbf{B}^{-1}\mathbf{E}=\mathbf{I},\label{eq:zf_condition}
\end{equation}
with $\mathbf{I}$ being $L\times L$ identity matrix. Condition \eqref{eq:zf_condition}
implies $\mathbf{G}\mathbf{H}\mathbf{F}\mathbf{B}^{-1}=\mathbf{E}^{-1}$
which leads to $\hat{\mathbf{y}}=\mathbf{a}-\mathbf{E}^{-1}\mathbf{d}+\mathbf{G}\mathbf{w}-\tilde{\mathbf{d}}.$
We obtain ZF property $\hat{\mathbf{y}}=\mathbf{a}+\mathbf{G}\mathbf{w}$
when 
\begin{equation}
\mathbf{E}^{-1}\mathbf{d}+\tilde{\mathbf{d}}=0\label{eq:modulo_condition}
\end{equation}
which is realized by a proper design of thresholds $\bm{\tau}$ and
$\bm{\tilde{\tau}}.$

\subsubsection{Modulo Threshold $\tau$}

\begin{figure}
\begin{centering}
\includegraphics[width=0.7\columnwidth]{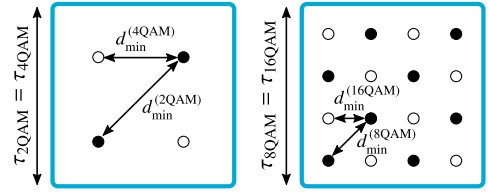}\vspace{-7pt}

\par\end{centering}

\protect\caption{Threshold $\tau$ for several considered square-shape QAM constellations.\label{fig:tau-treshold-for-qam}}
\vspace{-7pt}
\end{figure}

Size of $\tau$ is chosen to wrap constellations within $\tau\times\tau$
frame such that the distance from the edge point to the boundary is
half of minimal distance $d_{\min}.$ Figure~\ref{fig:tau-treshold-for-qam}
shows the frame for several QAM constellations considered in this
paper. It is straightforward to verify that square-shaped QAM constellations
including odd-bit cardinality variants (black points in Fig.\,\ref{fig:tau-treshold-for-qam})
have
$\tau=\sqrt{M}d_{\min}.$ Table~\ref{tab:Values-of-modulo} lists
values of $\tau$ considered here. 
\begin{table}
\vspace{-10pt}
\protect\caption{Threshold $\tau$ and energy increase $\Delta E$ due to modulo $\bm{\Gamma}_{\bm{\tau}}$
for considered unit-mean square-shape $M$-QAM constellations.\label{tab:Values-of-modulo}}
\vspace{-5pt}

\centering{}%
\begin{tabular}{|c|c|c|c|c|c|c|}
\hline 
$M$ & $2,4$ & $8,16$ & $32,64$ & $128,256$ & $512,1024$ & $2048,4096$\tabularnewline
\hline 
$\tau$ & 2.83 & 2.53 & 2.47 & 2.45 & 2.45 & 2.45\tabularnewline
\hline 
$\Delta E\,\mathrm{[dB]}$ & 1.25 & 0.28 & 0.068 & 0.017 & 0.0042 & 0.0011\tabularnewline
\hline 
\end{tabular}\vspace{-5pt}
\end{table}

\subsubsection{Per-Line Power Constraint}

Transmitted power on each line is constrained not to overcome a specified
limit. Without loss of generality, we use constellations normalized
to unit mean symbol energy $E\left[|a_{i}|^{2}\right]=1$ ($E[\star]$
denotes the statistical expectation) for which the energy limit implies
\begin{equation}
E\left[|x_{i}|^{2}\right]\leq1.\label{eq:energy_condition}
\end{equation}
We need to keep in mind and downscale constellations to pre-compensate
energy increase $\Delta E$ due to modulo $\bm{\Gamma}_{\bm{\tau}}.$
Square-shape even-bit QAM constellations have $\Delta E\simeq\nicefrac{M}{(M-1)}$~\cite{Ginis-Cioffi_2000}.
The same formula holds for square-shape odd-bit cardinalities if twice
higher value of $M$ is used. For example, $\Delta E$ is the same
for 2QAM and 4QAM as shown in Table~\ref{tab:Values-of-modulo}.
Odd-bit constellations with a square shape have significantly lower
$\Delta E$ than popular cross-shaped constellations~\cite{Muller-Lu-etal_2014}.

\subsection{Reference THP Scheme and Basic Properties\label{sub:Reference-Scheme-of}}

The proposed THP scheme enhances the scheme described in~\cite{Ginis-Cioffi_2000}
by ordered QR decomposition. Reference scheme~\cite{Ginis-Cioffi_2000}
is described by definition of the blocks in Fig.\,\ref{fig:Linearized-general-THP}
as 
\begin{equation}
\mathbf{E}=\mathbf{I},\quad\mathbf{B}=\mathrm{diag}(\mathbf{R})^{-1}\mathbf{R}^{H},\quad\mathbf{F}=\mathbf{Q},\quad\mathbf{G}=\mathrm{diag}(\mathbf{R})^{-1},\label{eq:thp_def}
\end{equation}
where $\mathrm{diag}(\mathbf{R})^{-1}=\mathrm{diag}\left(r_{11}^{-1},\dots r_{LL}^{-1}\right)$
with diagonal components $r_{ii}=\left[\mathbf{R}\right]_{ii}.$ Unitary
matrix $\mathbf{Q}$ and upper-triangular $\mathbf{R}$ follow from
QR decomposition of transposed channel matrix 
\begin{equation}
\mathbf{H}^{H}=\mathbf{Q}\mathbf{R}.\label{eq:qr_dec}
\end{equation}
Let us confirm basic properties introduced in Sec.\,\ref{sub:General-Scheme-of}.
The reference scheme fulfills ZF condition \eqref{eq:zf_condition}
\begin{equation}
\mathbf{G}\mathbf{H}\mathbf{F}\mathbf{B}^{-1}\mathbf{E}=\mathrm{diag}(\mathbf{R})^{-1}\mathbf{R}^{H}\mathbf{Q}^{H}\mathbf{Q}\mathbf{R}^{-H}\mathrm{diag}(\mathbf{R})\mathbf{I}=\mathbf{I}.
\end{equation}
Transmitted signal meets per-line energy constraint \eqref{eq:energy_condition}

\begin{equation}
E\left[|x_{i}|^{2}\right]=\sum_{j=1}^{L}|q_{ij}|^{2}E\left[|\tilde{x}_{j}|^{2}\right]=\sum_{j=1}^{L}|q_{ij}|^{2}=1,\label{eq:energy_check_thp}
\end{equation}
where $q_{ij}=\left[\mathbf{Q}\right]_{ij}.$ We use the fact that
$\mathbf{x}$ is approximately uncorrelated~\cite{Fischer-Windpassinger-etal_2002}
and the energy increase due to $\Gamma_{\bm{\tau}}$ has been pre-compensated
$E\left[|\tilde{x}_{j}|^{2}\right]=1.$ This last equality follows
from that unitary $\mathbf{Q}$ has unit-length rows. The decision
variable 
\begin{equation}
\hat{\mathbf{y}}=\mathbf{a}+\mathrm{diag}(\mathbf{R})^{-1}\mathbf{w}
\end{equation}
implies that output SNR at the \emph{i}th line is 
\begin{equation}
\gamma_{i}=\gamma_{base}\cdot r_{ii}^{2},\label{eq:snr_cioffi}
\end{equation}
where $\gamma_{base}$ is the baseline input SNR. The main diagonal
components $\left\{ r_{ii}^{2}\right\} _{i=1}^{L}$ can attain different
values providing different SNR at the each line. In this case, different
bit-loading per-line is required as well as different modulo threshold
$\tau_{i}\not=\tau_{j}$ in $\bm{\tau}=[\tau_{1},\dots,\tau_{L}]^{T},$
see Table~\ref{tab:Values-of-modulo} for actual values. There is
no integer precoding operation as $\mathbf{E}=\mathbf{I}$ and so
selection of $\bm{\tilde{\tau}}=\bm{\tau}$ fulfills modulo condition
\eqref{eq:modulo_condition}.

\section{Ordered Tomlinson-Harashima Precoding\label{sec:Proposed-Ordered-THP}}

\subsection{Ordered QR Decomposition}

The ordered THP scheme proposed here incorporates the ordered QR decomposition
of transposed channel
\begin{equation}
\mathbf{H}^{H}=\mathbf{Q}\mathbf{R}\mathbf{P}^{T}\label{eq:ordered_qr}
\end{equation}
into the reference scheme (Sec.\,\ref{sub:Reference-Scheme-of}).
Permutation matrix $\mathbf{P}$ describes arbitrary permutation $[1,\dots,L]^{T}\rightarrow[p_{1},\dots,p_{L}]^{T}$
as 
\begin{equation}
\mathbf{P}\left[\begin{array}{c}
1\\
\vdots\\
L
\end{array}\right]=\left[\begin{array}{c}
p_{1}\\
\vdots\\
p_{L}
\end{array}\right],\quad\mathrm{where}\quad\mathbf{P}=\left[\begin{array}{c}
\mathbf{e}_{p_{1}}\\
\vdots\\
\mathbf{e}_{p_{L}}
\end{array}\right]\label{eq:per_def}
\end{equation}
and $\mathbf{e}_{i}$ denotes a row vector with 1 in the \emph{i}th
position and 0 elsewhere. Note that $\mathbf{P}\mathbf{X}$ denotes
permutation of rows of $\mathbf{X}$ and $\mathbf{X}\mathbf{P}^{T}$
permutation of columns since $\mathbf{X}\mathbf{P}^{T}=\left(\mathbf{P}\mathbf{X}^{T}\right)^{T}.$

\subsection{Proposed THP Scheme and Basic Properties}

The ordered THP scheme is given by the following matrices 
\begin{equation}
\mathbf{E}=\mathbf{P}^{T},\mathbf{B}=\mathrm{diag}(\mathbf{R})^{-1}\mathbf{R}^{H},\,\mathbf{F}=\mathbf{Q},\,\mathbf{G}=\mathbf{P}\,\mathrm{diag}(\mathbf{R})^{-1}\mathbf{P}^{T}.\label{eq:oredered_thp_def}
\end{equation}
Key observation is that if diagonal matrix has permuted rows and columns
by the same permutation (as $\mathbf{G}$ in \eqref{eq:oredered_thp_def})
then it remains diagonal and can be performed by non-cooperating receivers.
Now, we show that ZF condition~\eqref{eq:zf_condition} is satisfied
\[
\mathbf{G}\mathbf{H}\mathbf{F}\mathbf{B}^{-1}\mathbf{E}=\mathbf{P}\mathrm{diag}(\mathbf{R})^{-1}\mathbf{P}^{T}\mathbf{P}\mathbf{R}^{H}\mathbf{Q}^{H}\mathbf{Q}\mathbf{R}^{-H}\mathrm{diag}(\mathbf{R})\mathbf{P}^{T}=\mathbf{I},
\]
since $\mathbf{P}^{T}$ describes inverse permutation and so $\mathbf{P}\mathbf{P}^{T}=\mathbf{I}.$
As in the reference scheme in Sec.\,\ref{sub:Reference-Scheme-of},
feedforward matrix $\mathbf{F}$ is unitary and therefore transmitted
signal meets per-line energy constraint~\eqref{eq:energy_condition}.
The decision variable $\hat{\mathbf{y}}=\mathbf{a}+\mathbf{P}\,\mathrm{diag}(\mathbf{R})^{-1}\mathbf{P}^{T}\mathbf{w}$
implies output SNR at the \emph{i}th line to be 

\begin{equation}
\gamma_{i}=\gamma_{base}\cdot r_{p_{i}p_{i}}^{2},\label{eq:snr_ordered_thp}
\end{equation}
where $\gamma_{base}$ denotes baseline SNR and $p_{i}$ is the\emph{
i}th element of permutation output \eqref{eq:per_def}. Similarly
to the reference THP, different values of main diagonal components
$\left\{ r_{p_{i}p_{i}}^{2}\right\} _{i=1}^{L}$ require different
bit-loading with thresholds $\bm{\tau}=[\tau_{1},\dots,\tau_{L}]^{T}$
where generally $\tau_{i}\not=\tau_{j}.$ Vector of thresholds $\tilde{\bm{\tau}}$
needs to be chosen to fulfill condition \eqref{eq:modulo_condition}
which means $\tilde{\bm{\tau}}=\mathbf{P}\bm{\tau}=[\tau_{p_{1}},\dots,\tau_{p_{L}}]^{T}.$ 
\begin{rem}
Any type of ordering $\mathbf{P}$ can be concatenated with the proposed
ordered THP and it is a degree of freedom to be exploited. The reference
scheme is obtained for the ordering $\mathbf{P}=\mathbf{I}$ which
means that optimized ordering can only improve the performance of
the reference scheme.
\end{rem}

\section{Optimized Ordering of THP in G.fast Downstream\label{sec:Optimized-Ordering-of}}

There is a rich number of possible orderings (see~\cite{Zu-Lamare-Haardt_2014}
and references therein) to be concatenated with the scheme proposed
here. Generally, different orderings lead to different SNR at each
line (\ref{eq:snr_ordered_thp}). Optimal selection is a multi-objective
optimization problem where utility target considering fairness has
significant impact on the result. We mainly focus on max-min fairness
by maximizing the minimum rate and thus provide the same quality of
service to each CPEs, although we discus sum-rate and simple combination
of both as well.

\subsection{V-BLAST (VB) Ordering}

The ordering strategy introduced in~\cite{Wolniansky-Foschini-etal_1998}
is the optimal max-min fair ordering maximizing the minimum SNR. The
algorithm requires multiple calculations of channel matrix pseudo-inverse
and so its complexity is much higher than the complexity of closely-related
semi-optimal algorithm~\cite{Wubben-Bohnke-etal_2001}. Instead of~\cite{Wolniansky-Foschini-etal_1998},
we pragmatically use~\cite{Wubben-Bohnke-etal_2001} since it performs
close to the optimum without the computational burdens. 

VB ordering~\cite{Wubben-Bohnke-etal_2001} is based on Gram-Schmidt
(GS) QR decomposition of the transposed channel $\mathbf{H}^{H}=\tilde{\mathbf{H}}=\mathbf{Q}\mathbf{R}.$
In the \emph{i}th iteration, the algorithm choses the column vector
$\tilde{\mathbf{h}}_{i}$ of $\tilde{\mathbf{H}}=\left[\tilde{\mathbf{h}}_{1},\dots,\tilde{\mathbf{h}}_{L}\right]$
which minimizes the diagonal element $r_{ii}=\left[\mathbf{R}\right]_{ii}$
given as 
\begin{equation}
r_{ii}=\left\Vert \tilde{\mathbf{h}}_{i}-\sum_{j=1}^{i-1}\left\langle \tilde{\mathbf{h}}_{i},\mathbf{q}_{j}\right\rangle \mathbf{q}_{j}\right\Vert ,\label{eq:rii}
\end{equation}
where $\left\langle \mathbf{h},\mathbf{q}\right\rangle =\mathbf{q}^{H}\mathbf{h}$
denotes an inner product. The order in which $\tilde{\mathbf{h}}$
are chosen forms permutation matrix $\mathbf{P}$ in (\ref{eq:ordered_qr}).
This strategy (``weakest first'') leads to the ordering which maximize
the minimum of $\left\{ r_{ii}^{2}\right\} _{i=1}^{L}$ elements and
so SNR~(\ref{eq:snr_ordered_thp}).

\subsection{Inverse V-BLAST (IVB) Ordering}

IVB describes ordering with opposite approach than VB. In each GS
iteration, always such a column vector $\tilde{\mathbf{h}}_{i}$ is
chosen for which diagonal element (\ref{eq:rii}) is maximal. It is
a greedy maximization approach which maximize the sum-rate. We freely
interchange sum-rate and mean-rate since the difference is just a
scaling factor. The IVB ordering (``strongest first'') is also known
as QR decomposition with pivoting~\cite{Meyer_MatrixAnalysis}.

\subsection{Dynamic Ordering (DO)}

\begin{figure}
\begin{centering}
\includegraphics[width=1\columnwidth]{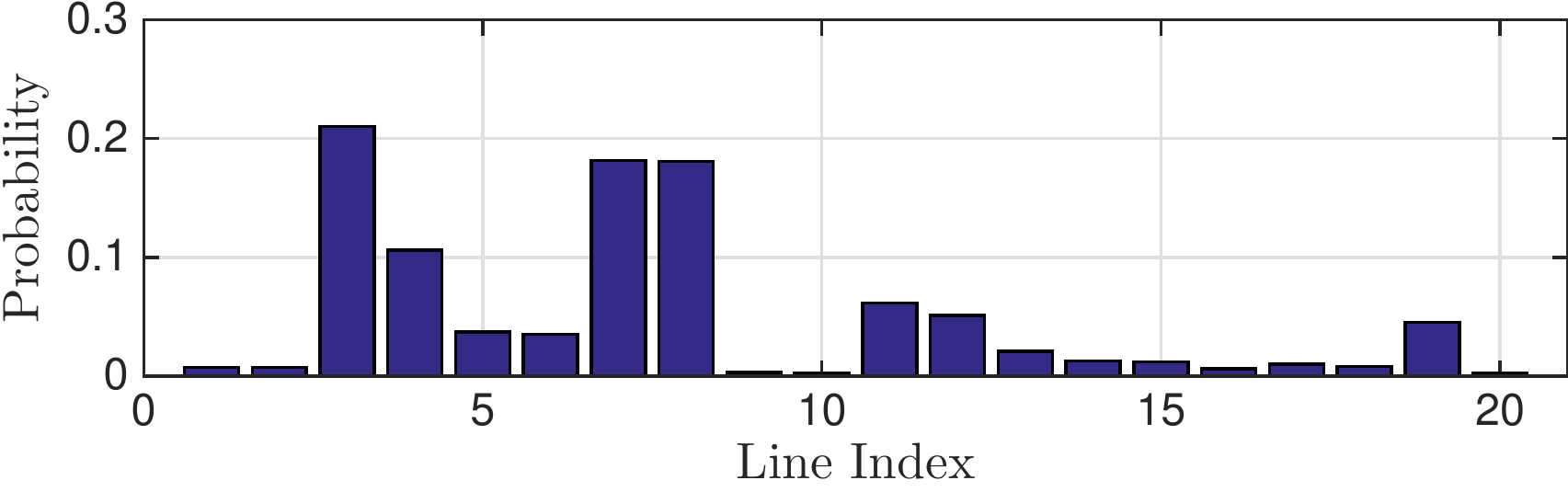}\vspace{-7pt}

\par\end{centering}

\protect\caption{Empirical probability density function of being weakest (i.e., being
selected as the first to enter GS procedure in V-BLAST ordering).\label{fig:Empirical-probability-density}}
\vspace{-7pt}
\end{figure}

Although VB is the optimal max-min fair ordering, we do not obtain
equal rates when aggregated over multiple DMT tones in numerical results
in Fig.\,\ref{fig:zf_rates}. So, if on a single tone, we selected
instead of VB ordering a different ordering in favor of the line with
the minimal aggregated rate, we would obtain the higher minimum. It
means that VB ordering is optimal on a single tone, but it does not
reach the global optimum if applied independently on each tones. The
reason behind is that G.fast channel does not have the same statistical
properties on each line. Some lines are more often the weakest lines
(being selected first by VB) due to asymmetric physical arrangement
of twisted-copper pairs within the cable bundle as confirmed by numerical
evaluation in Fig.~\ref{fig:Empirical-probability-density}.

\begin{figure}
\begin{centering}
\includegraphics[width=1\columnwidth]{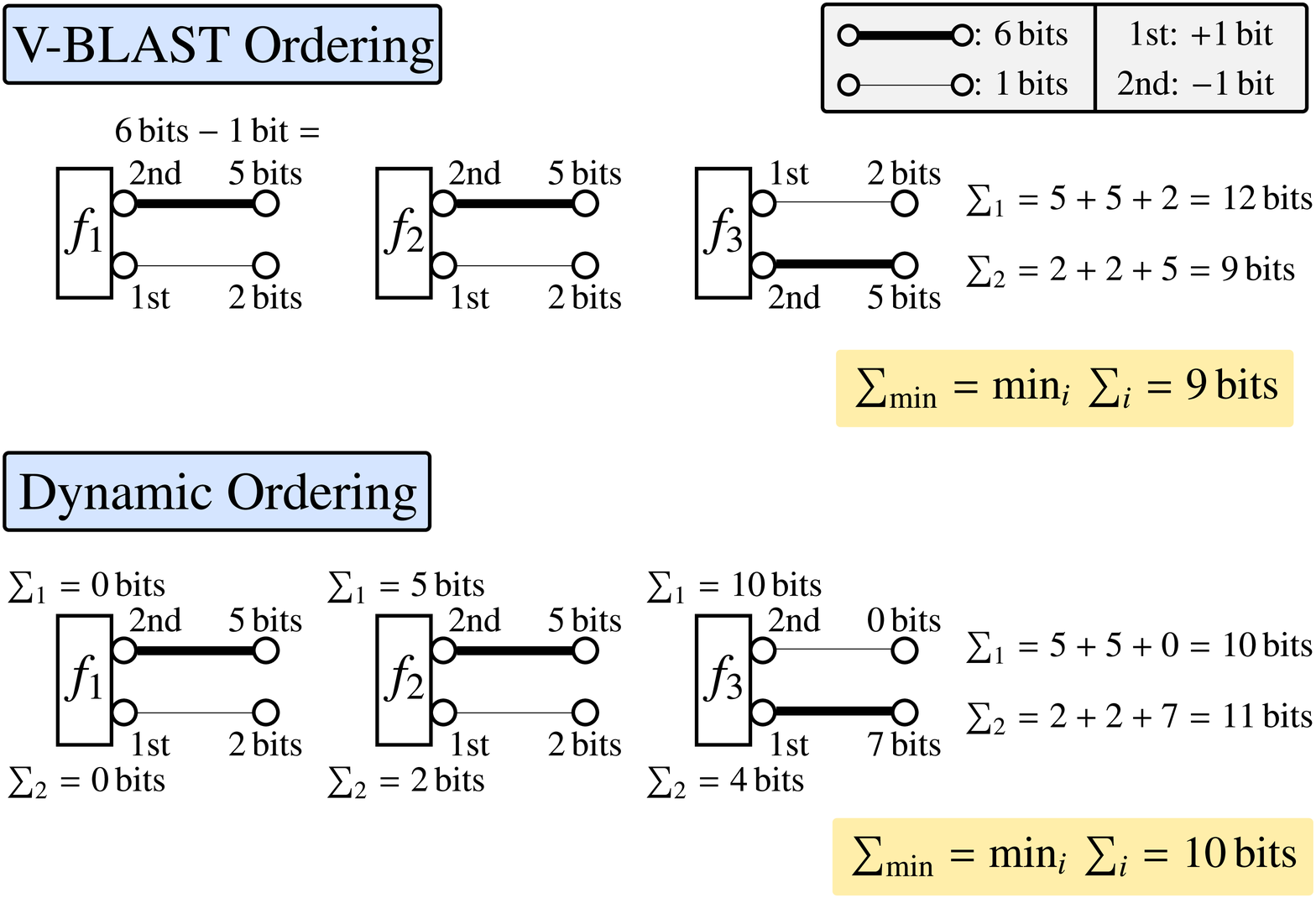}\vspace{-7pt}

\par\end{centering}

\protect\caption{Let us assume a DMT system with 2 lines and 3 tones $\left\{ f_{i}\right\} _{i=1}^{3}$.
Thick are strong lines with throughput of 6\,bits and thin are weak
lines with throughput of 1\,bit. Entering GS procedure first is an
advantage increasing the throughput by $+1\,\mathrm{bit}$ and second
decreasing the throughput by $-1\,\mathrm{bit}$. The minimum aggregated
rate $\sum_{\min}$ of V-BLAST (``weakest first'') is lower than
DO (``aggregated minimum first''). Symbol $\sum_{i}$ denotes aggregated
rate at the \emph{i}th line. We assume initial order at $f_{1}$ of
DO to be given by VB.\label{fig:vb-do-example}}
\vspace{-7pt}
\end{figure}

We propose DO strategy taking into account this statistical asymmetry
providing the highest minimum aggregated rate. DO ordering is inspired
by VB approach which states that being taken first into GS is an advantage.
Instead of VB ``weakest first'' approach applied independently on
each tone, we propose to take first the line with so far minimum aggregated
rate (``aggregated minimum first''). DO is ordering with memory
deciding the order inductively in sequence. If the ordering on tone
index 1 to $i-1$ has been already chosen, then DO orders the lines
on the \emph{i}th tone as the order of bit-loading aggregated over
tones from 1 to $i-1.$ Figure~\ref{fig:vb-do-example} shows an
illustrative example explaining why DO provides higher aggregated
minimum than VB.
Complexity of DO is negligible, since the order is given by cumulative
summation performed once at the beginning of transmission. The order
is computed outside of QR algorithm and can be connected to whatever
type of QR implementation, not only the one based on GS as in~\cite{Wubben-Bohnke-etal_2001}.

\subsection{Frequency-Sharing Between DO and IVB}

\begin{figure}
\begin{centering}
\includegraphics[width=0.8\columnwidth]{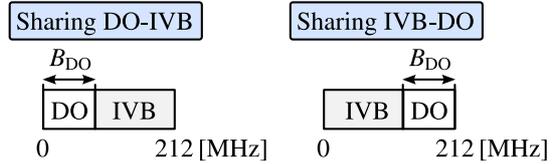}\vspace{-7pt}

\par\end{centering}

\protect\caption{Frequency sharing between two extrema types of orderings DO (maximizing
minimum) and IVB (maximizing sum-rate). Parameter $B_{\mathrm{DO}}$
describes bandwidth assigned to DO. \label{fig:Frequency-sharing-between-two}}
\vspace{-7pt}
\end{figure}

\begin{figure}
\begin{centering}
\includegraphics[width=1\columnwidth]{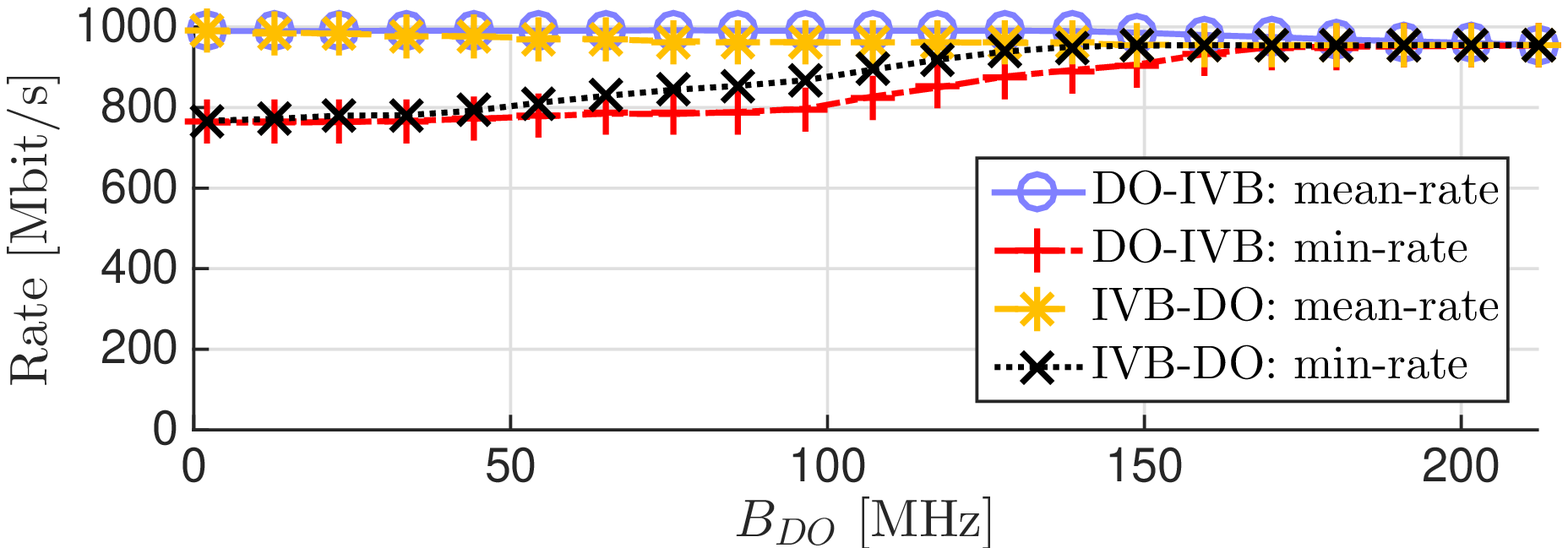}\vspace{-7pt}

\par\end{centering}

\protect\caption{Frequency sharing between DO and IVB enables to adjust trade off between
mean and min-rate. For instance, both min-rate and mean-rate equal
to $\sim950\,\mathrm{Mbps}$ for DO-IVB sharing with $B_{\mathrm{DO}}=212\,\mathrm{MHz}$,
but if we permit a slight decrease of minimum, we could have mean-rate
$\sim975\,\mathrm{Mbps}$ while having minimum still $\sim950\,\mathrm{Mbps}$
(here $B_{\mathrm{DO}}\simeq170\,\mathrm{MHz}$). Similarly, when
mean-rate is priority, $B_{\mathrm{DO}}\simeq125\,\mathrm{MHz}$ increases
min-rate from $\sim760\,\mathrm{Mbps}$ ($B_{\mathrm{DO}}\simeq0\,\mathrm{MHz}$)
to $\sim875\,\mathrm{Mbps}$ keeping the same mean-rate.\label{fig:Mean-rate-and}}
\vspace{-7pt}
\end{figure}

IVB ordering maximizes sum-rate on a single tone as well as when applied
independently on multiple tones, unlike in max-min case of VB and
DO ordering. We propose a simple frequency-sharing between two extrema
types of ordering IVB and DO to adjust the fairness among CPEs. By
frequency-sharing, we mean similar concept as time-sharing, but in
the frequency domain. We propose to divide bandwidth on lower and
upper parts where we expect different behavior (e.g., diagonal dominant
property is present only on lower frequencies as shown in Fig.\,\ref{fig:Diagonal-dominant-property}).
DO ordering is allocated to lower frequencies in case DO-IVB and to
higher frequencies in case IVB-DO as shown in Fig.\,\ref{fig:Frequency-sharing-between-two}.
Numerical evaluation in Fig.\,\ref{fig:Mean-rate-and} shows that
DO-IVB sharing achieves better results. 

\section{Comparison with Ordered Equal Rate THP (ER-THP)\label{sec:Comparison-with-Ordered}}

Ordering optimization has been introduced in~\cite{Fischer-Windpassinger-etal_2002},~\cite{Joham-Brehmer-Utschick_2004}
for THP structure which we denote as ER-THP (term centralized THP
is also used \cite{Huang-Zhou-Wang_2008}). Label 'equal rate' corresponds
to the feature that ER-THP provides constant SNR. The proposed THP
scheme and ER-THP have essentially different structure.

\subsection{Ordered ER-THP Scheme and Basic Properties}

Ordered ER-THP is defined by the following matrices

\begin{equation}
\mathbf{E}=\mathbf{P}^{T},\mathbf{B}=\mathbf{R}^{H}\mathrm{diag}(\mathbf{R})^{-1},\mathbf{F}=\frac{1}{g}\,\mathbf{Q}\,\mathrm{diag}(\mathbf{R})^{-1},\,\mathbf{G}=g\mathbf{I},\label{eq:er_thp}
\end{equation}
where ordered QR decomposition (\ref{eq:ordered_qr}) is used. Automatic
gain control scaling $g$ establishes power constrain (\ref{eq:energy_condition})
so 
\begin{equation}
E\left[|x_{i}|^{2}\right]=\frac{1}{g^{2}}\sum_{j=1}^{L}\left|\tilde{f}_{ij}\right|^{2}\leq1,
\end{equation}
with labeling $\tilde{\mathbf{F}}=\mathbf{Q}\,\mathrm{diag}(\mathbf{R})^{-1}$
and $\left[\tilde{\mathbf{F}}\right]_{ij}=\tilde{f}_{ij}.$ The constraint
is fulfilled by the following scaling using $\left(2,\infty\right)$-mixed
norm $\left\Vert \star\right\Vert _{2,\infty}$ as 
\begin{equation}
g^{2}=\left\Vert \tilde{\mathbf{F}}^{T}\right\Vert _{2,\infty}^{2}\triangleq\max_{i}\sum_{j=1}^{L}\left|\tilde{f}_{ij}\right|^{2}=\max_{i}\sum_{j=1}^{L}\nicefrac{\left|q_{ij}\right|^{2}}{r_{jj}^{2}}.\label{eq:mix_norm}
\end{equation}
Notice, that average power constrain assumed in~\cite{Fischer-Windpassinger-etal_2002},~\cite{Joham-Brehmer-Utschick_2004}
leads to the scaling with Frobenius norm $\left\Vert \star\right\Vert _{F}$
as $g^{2}=\nicefrac{1}{L}\left\Vert \tilde{\mathbf{F}}\right\Vert _{F}^{2}\triangleq\nicefrac{1}{L}\,\mathrm{tr}\left(\tilde{\mathbf{F}}\tilde{\mathbf{F}}^{H}\right).$
We confirm that ER-THP meets ZF condition \eqref{eq:zf_condition}

\[
\mathbf{G}\mathbf{H}\mathbf{F}\mathbf{B}^{-1}\mathbf{E}=g\mathbf{I}\mathbf{P}\mathbf{R}^{H}\mathbf{Q}^{H}\frac{1}{g}\mathbf{Q}\,\mathrm{diag}(\mathbf{R})^{-1}\mathrm{diag}(\mathbf{R})\mathbf{R}^{-H}\mathbf{P}^{T}=\mathbf{I}.
\]
The decision variable $\hat{\mathbf{y}}=\mathbf{a}+g\mathbf{w}$ implies
output SNR to be 

\begin{equation}
\gamma_{i}=\gamma_{base}\cdot\nicefrac{1}{g^{2}},\label{eq:snr_vb_fischer}
\end{equation}
where $\gamma_{base}$ denotes baseline SNR. Constant SNR yields the
same bit-loading and the same modulo threshold $\bm{\tau}=[\tau,\dots,\tau]^{T}$
on every line, therefore vector of thresholds $\tilde{\bm{\tau}}=\bm{\tau}$
fulfills modulo condition \eqref{eq:modulo_condition}.

\subsection{VB Ordered and Lattice Reduced (LR) ER-THP}

Performance of ER-THP is given by scaling factor $g^{2}.$ Using inequality
$\nicefrac{1}{r_{ii}^{2}}\leq\nicefrac{1}{\min_{i}r_{ii}^{2}}$ ,
we rephrase~(\ref{eq:mix_norm}) as 
\begin{equation}
g^{2}=\max_{i}\sum_{j=1}^{L}\frac{\left|q_{ij}\right|^{2}}{r_{jj}^{2}}\leq\max_{i}\sum_{j=1}^{L}\frac{\left|q_{ij}\right|^{2}}{\min_{k}r_{kk}^{2}}=\frac{1}{\min_{k}r_{kk}^{2}}.
\end{equation}
We see that $g^{2}$ is minimized when $\min_{k}r_{kk}^{2}$ is as
large as possible, therefore VB ordering (maximizing the minimum of
$\left\{ r_{ii}\right\} _{i=1}^{L}$) is again preferable. Reference~\cite{Fischer-Windpassinger_2003}
shows that even smaller value of $g^{2}$ is obtained with LR QR decomposition
\begin{equation}
\mathbf{H}^{H}=\mathbf{Q}\mathbf{R}\mathbf{T}^{-1},\label{eq:lrvb_qr}
\end{equation}
where QR decomposes reduced channel as $\tilde{\mathbf{H}}=\mathbf{Q}\mathbf{R}$
where reduced channel is $\tilde{\mathbf{H}}=\mathbf{H}^{H}\mathbf{T}$
and $\mathbf{T}$ is a unimodular integer matrix. LR ER-THP is given
by (\ref{eq:er_thp}) using decomposition (\ref{eq:lrvb_qr}) where
$\mathbf{E}=\mathbf{T}^{H}$. We use familiar LLL implementation of
LR with moderate algorithm complexity parameter $\delta=\nicefrac{3}{4}$~\cite{Wubben-Seethaler-etal_2011}. 

Unfortunately, LR decomposition (\ref{eq:lrvb_qr}) cannot be used
in the proposed THP scheme (\ref{eq:oredered_thp_def}) with $\mathbf{E}=\mathbf{T}^{H},$
as matrix $\mathbf{G}=\mathbf{T}^{-H}\,\mathrm{diag}(\mathbf{R})^{-1}\mathbf{T}^{H}$
(unlike in the case of ordering) is not diagonal anymore and thus
cannot be performed by non-cooperating CPEs. For the sake of comparison,
we consider ER-THP scheme enhanced by both LR and VB ordering with
complexity parameter set to an extreme value $\delta=1.$ The scheme
has impractical implementation complexity but gives the highest SNR
as confirmed by simulations in Fig.\,\ref{fig:zf_rates}.

\section{Performance Evaluation over G.fast Channel \label{sec:Performance-Evaluation}}

\subsection{Tested 100m Long Paper-Insulated Cable}

\begin{figure}
\begin{centering}
\includegraphics[width=1\columnwidth]{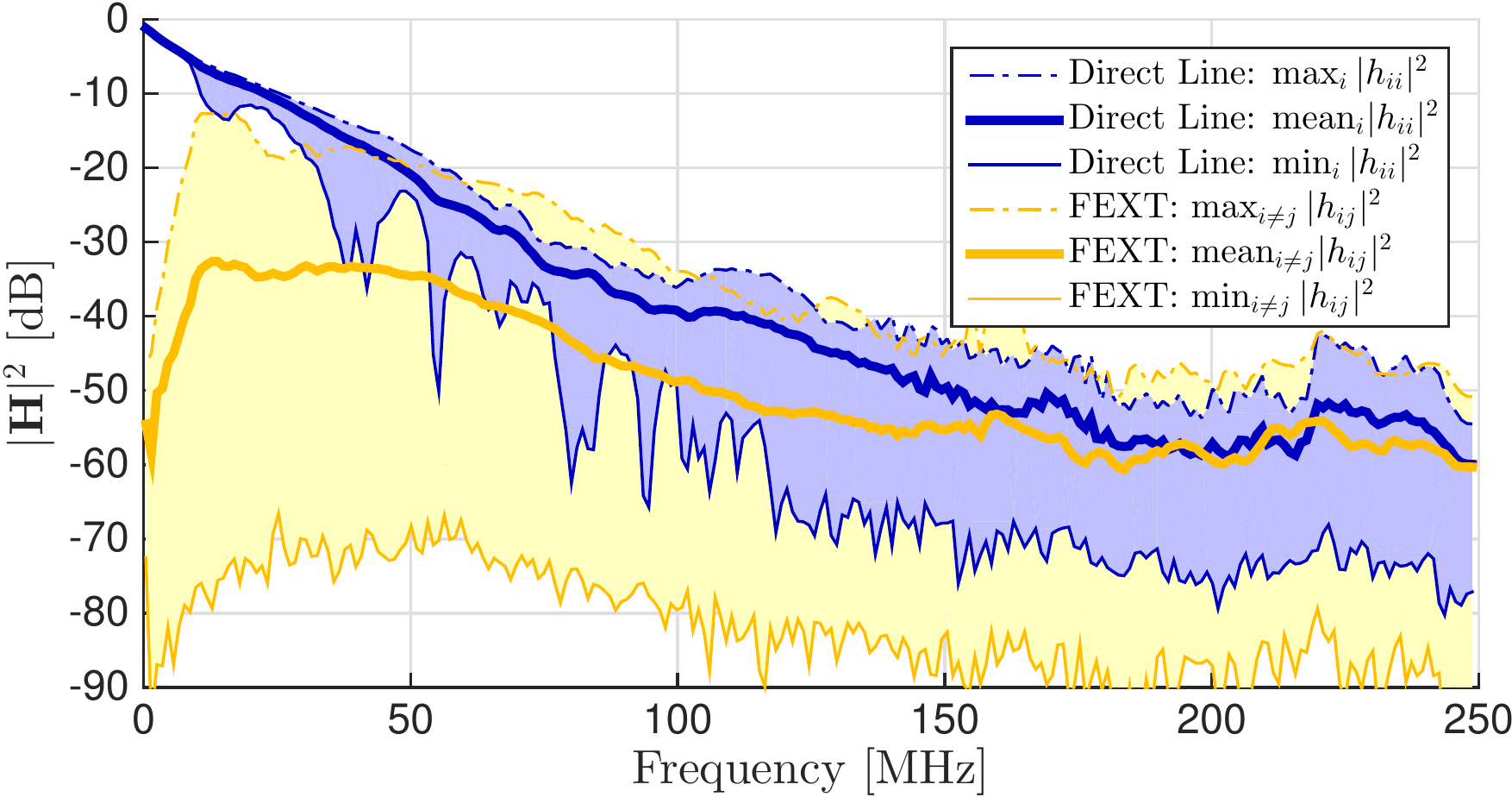}\vspace{-7pt}

\par\end{centering}

\protect\caption{Direct line and crosstalk power characteristics of tested 100\,m
long paper-insulated cable. Values are smoothen by average over $1\,\mathrm{MHz}$
bin. \label{fig:Direct-line-and}}
\vspace{-7pt}
\end{figure}

\begin{figure}
\begin{centering}
\includegraphics[width=0.5\columnwidth]{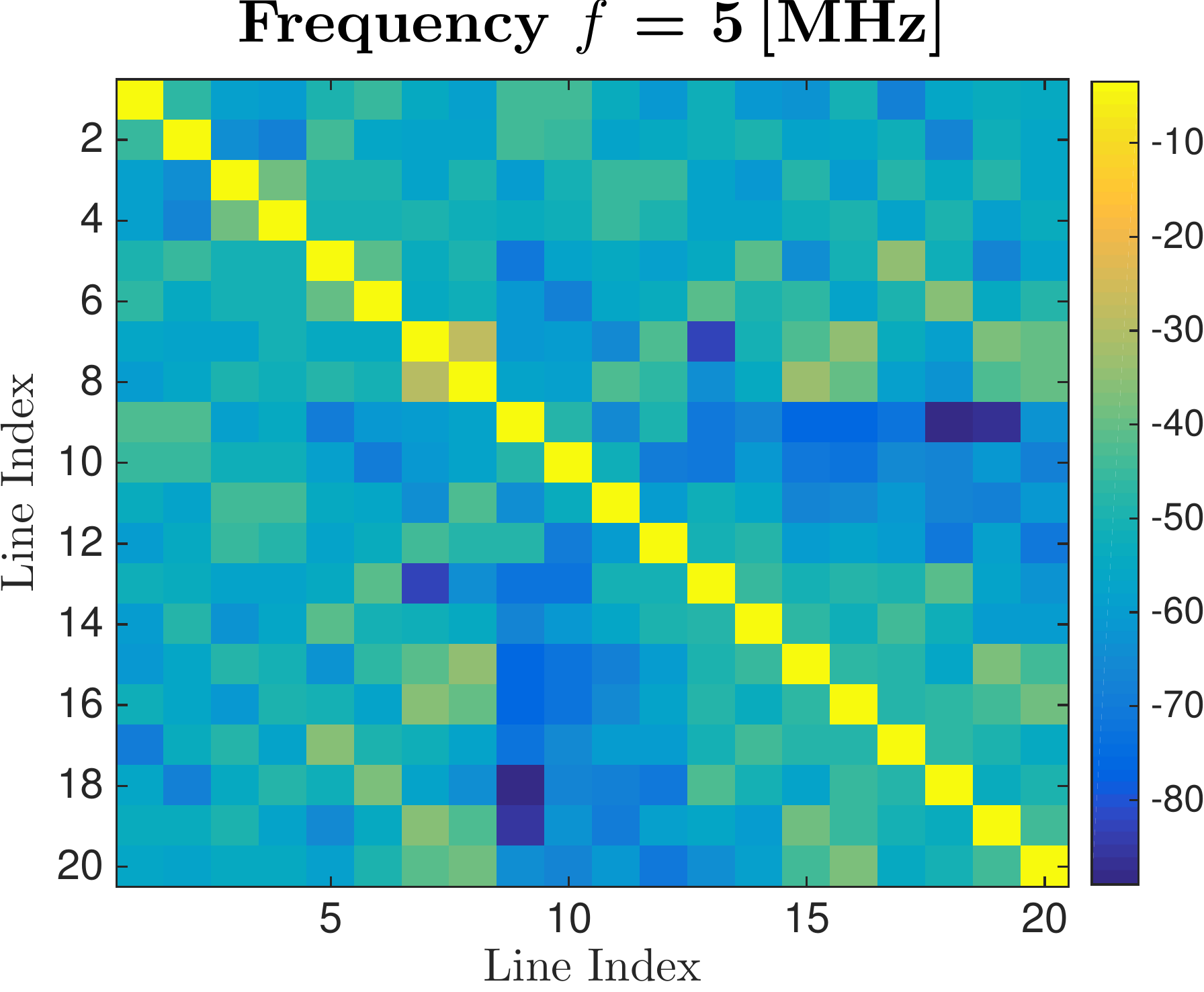}~\includegraphics[width=0.5\columnwidth]{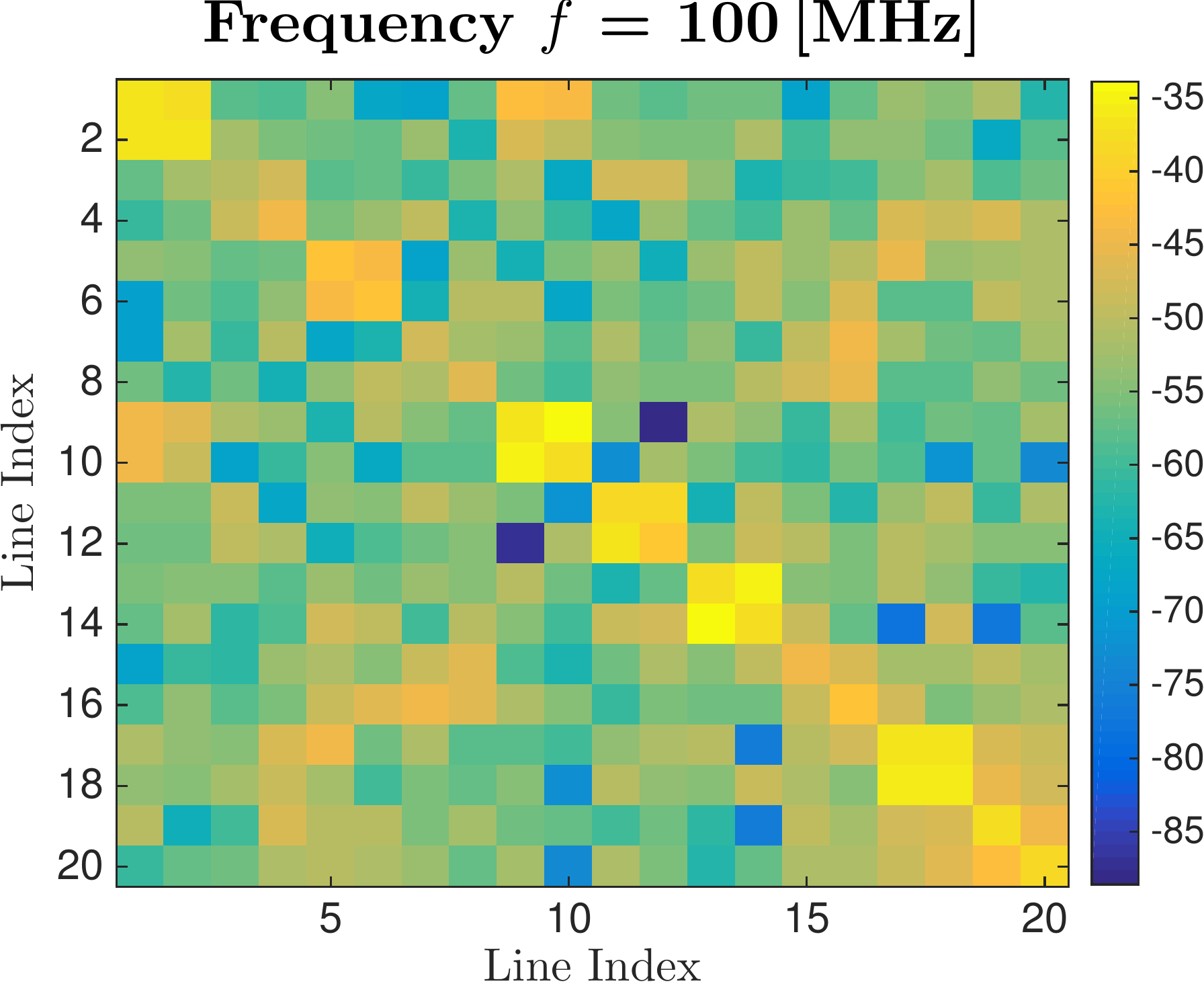}\vspace{-7pt}

\par\end{centering}

\protect\caption{Diagonal dominant property of channel matrix $\left|\mathbf{H}\right|^{2}\,\left[\mathrm{dB}\right]$
at carrier frequency 5\,{[}MHZ{]} is not present at higher frequency
100\,{[}MHz{]}.\label{fig:Diagonal-dominant-property}}
\vspace{-7pt}
\end{figure}

Figures~\ref{fig:Direct-line-and} and \ref{fig:Diagonal-dominant-property}
show strong FEXT of 100m long paper-insulated G.fast cable where diagonal
dominant property of VDSL2 is not present any more.

\subsection{Evaluation Procedure and Simulation Parameters}

Bit-loading at the \emph{i}th line is computed according to \cite{Adhoc-Convenor_2012-ITU-T}
by insertion of SNR $\gamma_{i}$ (\ref{eq:snr_cioffi}), (\ref{eq:snr_ordered_thp}),
(\ref{eq:snr_vb_fischer}) into the gap formula 
\begin{equation}
b_{i}=\left\lfloor \log_{2}\left(1+\nicefrac{\gamma_{i}}{\Gamma}\right)\right\rfloor ,\quad\mathrm{for}\quad b_{i}\in[2,12]\label{eq:gap}
\end{equation}
and $b_{i}=0$ otherwise, where symbol $\left\lfloor \star\right\rfloor $
denotes floor operation and gap $\Gamma=\mathrm{Shannon\,gap}+\mathrm{margin}-\mathrm{coding\,gain}$
{[}dB{]}. Aggregated rates
\begin{figure}
\begin{centering}
\vspace{-5pt}
\includegraphics[width=1.05\columnwidth]{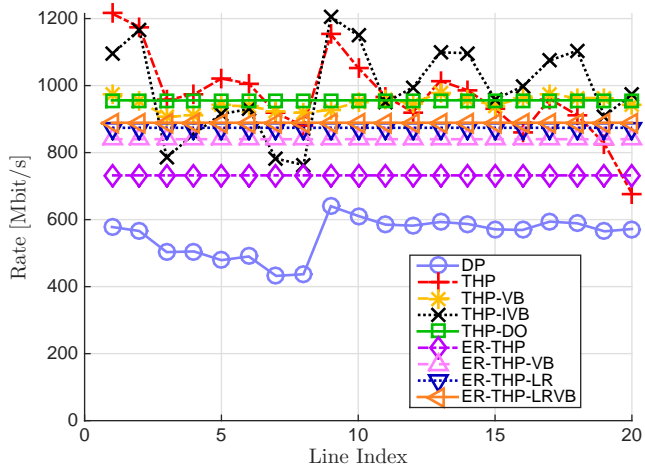}\vspace{-7pt}

\par\end{centering}

\protect\caption{Aggregated rates of considered precoding schemes over tested 100m
long paper-insulated cable. The acronyms and basic rate statistics
of considered precoding schemes are summarized in Table~\ref{tab:Mean-and-minimum}.\label{fig:zf_rates}}
\vspace{-7pt}
\end{figure}
 are obtained by summation of bit-loading (\ref{eq:gap}) over all
DMT tones multiplied by $\Delta f\,\left(1-\mathrm{Framing\,Overhead}\right)$,
where $\Delta f$ denotes the tone spacing. Considered parameters
are listed in Table~\ref{tab:Simulation-Parameters}.
\begin{table}
\protect\caption{Simulation parameters \cite{Adhoc-Convenor_2012-ITU-T}\label{tab:Simulation-Parameters}}
\vspace{-7pt}

\noindent \centering{}{\scriptsize{}}%
\begin{tabular}{|c|c||c|c|}
\hline 
{\scriptsize{}Transmit PSD} & {\scriptsize{}$-76\,\mathrm{[dBm/Hz]}$} & {\scriptsize{}Coding Gain} & {\scriptsize{}$5\,\mathrm{[dB]}$}\tabularnewline
{\scriptsize{}Noise PSD} & {\scriptsize{}$-140\,\mathrm{[dBm/Hz]}$} & {\scriptsize{}Shannon Gap} & {\scriptsize{}$9.8\,\mathrm{[dB]}$}\tabularnewline
{\scriptsize{}Band} & {\scriptsize{}$2.1-212\,\mathrm{[MHz]}$} & {\scriptsize{}Bit Loading} & {\scriptsize{}$2-12\,\mathrm{[bits]}$}\tabularnewline
{\scriptsize{}Tone Spacing $\Delta f$} & {\scriptsize{}$51.750\,\mathrm{[kHz]}$} & {\scriptsize{}Framing Overhead} & {\scriptsize{}$12\,\%$}\tabularnewline
{\scriptsize{}Margin} & {\scriptsize{}$6\,\mathrm{[dB]}$} &  & \tabularnewline
\hline 
\end{tabular}\vspace{-7pt}
\end{table}
We use bit-loading modification described in~\cite{Muller-Lu-etal_2014}
to incorporate energy increases $\Delta E$ (shown in Table~\ref{tab:Values-of-modulo})
due to modulo $\bm{\Gamma}_{\bm{\tau}}$. The algorithm allocates
bits according to (\ref{eq:gap}) and then recomputes SNR corrected
by $\Delta E$ and update bit-allocation accordingly.

\subsection{Numerical Results}

Numerical results in Fig.\,\ref{fig:zf_rates} and Table~\ref{tab:Mean-and-minimum}
compare several FEXT cancellation methods in DS. Non-linear precoding
based on THP or ER-THP clearly outperforms linear DP precoding used
in VDSL2~\cite{Cendrillon-Moonen-etal_2004}. We confirm that ER-THP
provides constant aggregated rates to all users, where gains by VB
ordering and LR are significant. THP with un-equal rates provides
higher sum-rate than ER-THP. As expected, VB ordered THP considerably
increase min-rate and IVB ordered THP considerably increase sum-rate.
Proposed DO ordered THP provides the highest aggregated minimum rate
among all considered methods. The achieved rates are fairly stable
vs. line index being close to G.fast target of 1\,Gbps.
\begin{table}
\protect\caption{Mean and minimum aggregated rates of several FEXT cancellation precoding
schemes depicted in Fig.~\ref{fig:zf_rates}. The rates are in $\mathrm{Mbits}/\mathrm{s}$.\label{tab:Mean-and-minimum}}

\noindent \centering{}\vspace{-7pt}
{\footnotesize{}}%
\begin{tabular}{|c|c|c|}
\hline 
{\footnotesize{}Acronym} & {\footnotesize{}Precoding Scheme} & {\footnotesize{}{[}mean,min{]}-rate}\tabularnewline
\hline 
{\footnotesize{}DP} & {\footnotesize{}Diagonal Precoding} & {\footnotesize{}$[552,432]$ }\tabularnewline
{\footnotesize{}THP} & {\footnotesize{}Tomlinson-Harashima Precoding} & {\footnotesize{}$[970,678]$ }\tabularnewline
{\footnotesize{}THP-VB} & {\footnotesize{}THP using VB ordering} & {\footnotesize{}$[947,907]$ }\tabularnewline
{\footnotesize{}THP-IVB} & {\footnotesize{}THP using Inverse VB ordering} & {\footnotesize{}$[990,763]$}\tabularnewline
{\footnotesize{}THP-DO} & {\footnotesize{}THP using Dynamic Ordering} & {\footnotesize{}$[956,955]$ }\tabularnewline
{\footnotesize{}ER-THP} & {\footnotesize{}Equal-Rate THP} & {\footnotesize{}$[732,732]$ }\tabularnewline
{\footnotesize{}ER-THP-VB} & {\footnotesize{}ER-THP using VB ordering} & {\footnotesize{}$[840,840]$ }\tabularnewline
{\footnotesize{}ER-THP-LR} & {\footnotesize{}ER-THP using Lattice Reduction} & {\footnotesize{}$[874,874]$ }\tabularnewline
{\footnotesize{}ER-THP-LRVB} & {\footnotesize{}ER-THP-LR using VB ordering} & {\footnotesize{}$[889,889]$ }\tabularnewline
\hline 
\end{tabular}\vspace{-7pt}
\end{table}

\section{Conclusion\label{sec:Conclusion}}

Contribution of this paper is two fold, a new ordered THP scheme and
a novel Dynamic Ordering (DO) strategy has been proposed, which together
leads to the highest aggregated minimum rate in G.fast downstream.
Unlike existing ordered Equal Rate (ER) THP scheme, the proposed scheme
better adapts to asymmetric channel statistics of G.fast channel.
Although the results are related to concrete G.fast settings, the
proposed scheme has universal application in general multiple-input
multiple-output systems including wireless scenario (e.g., paper~\cite{Huang-Zhou-Wang_2008}
shows that sum-rate of THP is always higher or equal than sum-rate
of ER-THP when the same type of channel matrix decomposition is considered).

\bibliographystyle{IEEEtran}
\bibliography{lib/bib/papers,lib/bib/books,lib/bib/dsl}

\end{document}